\def\papertitle{Improving Unsupervised Clean-to-Rendered Guitar Tone Transformation Using GANs and Integrated Unaligned Clean Data}
\def\paperauthorA{Yu-Hua Chen}
\def\paperauthorB{Woosung Choi$^{*}$}
\def\paperauthorC{Wei-Hsiang Liao$^{*}$}
\def\paperauthorD{Marco A. Martínez-Ramírez$^{*}$}
\def\paperauthorE{Kin Wai Cheuk$^{*}$}
\def\paperauthorF{Yuki Mitsufuji$^{*,\dagger}$}
\def\paperauthorG{Jyh-Shing Roger Jang}
\def\paperauthorH{Yi-Hsuan Yang}
\documentclass[twoside,a4paper]{article}
\usepackage{etoolbox}
\usepackage{dafx_24}
\usepackage{amsmath,amssymb,amsfonts,amsthm}
\usepackage{euscript}
\usepackage[T1]{fontenc}
\usepackage[utf8]{inputenc}
\usepackage{ifpdf}
\usepackage[english]{babel}
\usepackage{caption}
\usepackage{subfigure} % or can use subcaption package
\usepackage{color}
\usepackage[table,xcdraw]{xcolor}
\usepackage{adjustbox}
\usepackage{multirow}
\usepackage{booktabs}
\usepackage{amsmath}
\input glyphtounicode
\ninept

\newcounter{numauth}
\newcounter{listcnt}
\newcommand\authcnt[1]{\ifdefined#1 \stepcounter{numauth} \fi}

\newcommand\addauth[1]{
\ifdefined#1 
\stepcounter{listcnt}
\ifnum \value{listcnt}<\value{numauth}
\appto\authorslist{, #1}
\else
\appto\authorslist{~and~#1}
\fi
\fi}
\authcnt{\paperauthorB}
\authcnt{\paperauthorC}
\authcnt{\paperauthorD}
\authcnt{\paperauthorE}
\authcnt{\paperauthorF}
\authcnt{\paperauthorG}
\authcnt{\paperauthorH}
\authcnt{\paperauthorI}
\authcnt{\paperauthorJ}
\def\authorslist{\paperauthorA}
\addauth{\paperauthorB}
\addauth{\paperauthorC}
\addauth{\paperauthorD}
\addauth{\paperauthorE}
\addauth{\paperauthorF}
\addauth{\paperauthorG}
\addauth{\paperauthorH}
\addauth{\paperauthorI}
\addauth{\paperauthorJ}
\usepackage{times}
\newif\ifpdf
\ifx\pdfoutput\relax
\else
   \ifcase\pdfoutput
      \pdffalse
   \else
      \pdftrue
\fi

\ifpdf % compiling with pdflatex
  \usepackage[pdftex,
    pdftitle={\papertitle},
    pdfauthor={\authorslist},
    pdfsubject={Proceedings of the 27th International Conference on Digital Audio Effects (DAFx24)},
    colorlinks=false, % links are activated as color boxes instead of color text
    bookmarksnumbered, % use section numbers with bookmarks
    pdfstartview=XYZ % start with zoom=100% instead of full screen; especially useful if working with a big screen :-)
  ]{hyperref}
\else % compiling with latex
  \usepackage[dvips,
    pdftitle={\papertitle},
    pdfauthor={\authorslist},
    pdfsubject={Proceedings of the 27th International Conference on Digital Audio Effects (DAFx24)},
    colorlinks=false, % no color links
    bookmarksnumbered, % use section numbers with bookmarks
    pdfstartview=XYZ % start with zoom=100% instead of full screen
  ]{hyperref}
\fi
\usepackage[hypcap=true]{caption}
\title{\papertitle}

\fouraffiliations{
\paperauthorA \,\thanks{\vspace{-3mm}}}
{Graduate Institute of Networking and Multimedia\\National Taiwan University\\Taipei, Taiwan\\{\tt \href{mailto:f08946011@ntu.edu.tw}{f08946011@ntu.edu.tw}}}
{\paperauthorB, \paperauthorC, \paperauthorD, \paperauthorE, \paperauthorF}
{$^{*}$ {Sony AI}, $^{\dagger}$ {Sony Group Corporation}\\Tokyo, Japan\\{\tt {woosung.choi@sony.com}}}
{\paperauthorG \,}
{{Graduate Institute of Networking and Multimedia}\\National Taiwan University\\Taipei, Taiwan \\{\tt \href{mailto:jang@mirlab.org}{jang@mirlab.org}}}
{\paperauthorH \,}
{{Department of Electrical Engineering}\\National Taiwan University\\Taipei, Taiwan\\{\tt \href{mailto:yhyangtw@ntu.edu.tw}{yhyangtw@ntu.edu.tw}}}
\begin{document}
% more pdf-tex settings:
\ifpdf % used graphic file format for pdflatex
  \DeclareGraphicsExtensions{.png,.jpg,.pdf}
\else  % used graphic file format for latex
  \DeclareGraphicsExtensions{.eps}
\fi
%\makeatletter
%\pdfbookmark[0]{\@pdftitle}{title}
%\makeatother
\newcommand{\yuhuarevised}[1]{{\color{black}#1}}
\newcommand{\ericrevised}[1]{{\color{black}#1}}
\newcommand{\woosungvised}[1]{{\color{black}#1}}
\maketitle
\begin{abstract} 
Recent years have seen increasing interest in applying deep learning methods to the modeling of guitar amplifiers or effect pedals. 
Existing methods are mainly based on the supervised approach, requiring temporally-aligned data pairs of unprocessed and rendered audio.
However, this approach does not scale well, 
due to the complicated process involved in creating the data pairs.
A very recent work done by Wright \emph{et al.} has explored the potential of leveraging unpaired data for training, using a generative adversarial network (GAN)-based framework. 
This paper extends their work by using more advanced discriminators in the GAN, and using more unpaired data for training. 
Specifically, drawing inspiration from recent advancements in neural vocoders, we employ in our GAN-based model for guitar amplifier modeling two sets of discriminators, one based on multi-scale discriminator (MSD) and the other multi-period discriminator (MPD). Moreover, we experiment with adding unprocessed audio signals that do not have the corresponding rendered audio of a target tone to the training data, to see how much the GAN model benefits from the unpaired data. Our experiments show that the proposed two extensions contribute to the modeling of both low-gain and high-gain guitar amplifiers.
\end{abstract}

% Eric comment
% If nam l1 l2 really 
% L1 and L2 perfromance acceptable
% we can do better

\section{Introduction}
\label{sec:intro}
Amplifier modeling involves developing algorithms to emulate the behavior of real amplifiers. The amplifiers typically discussed in the literature are vacuum tube amplifiers. This task can also be considered a virtual analog (VA) modeling problem. Recent studies have demonstrated the potential to apply neural networks to VA modeling tasks using supervised learning. Various network architectures have been proposed in the literature, such as convolution-based and recurrent-based networks  \cite{wright2020real,damskagg2019deep, steinmetz2021steerable,769f627fa4fe49569bd207f6b1d32dc3, juvela2023end}.

% The VA modeling approaches can be summarized in two folds, which are 'white box' and 'black box' method. The “white-box” VA technique creates a digital version of the device itself. In order to simulate every components in the device, it requires the analysis of the circuit scheme of the device. Since the result is accurate and fully controllable which make is a popular method that many plugin companies to design their own product based on a released circuit scheme. Contrary, the "black-box" method can be developed without having a background knowledge of target device itself. However, these techniques also bring same flaws such as uncontrollable and difficult to optimizing. 

% Even though these approaches are feasible for having either a DSP-based model or neural network as a VA modeling model, especially in guitar amplifier modeling task. They all required a paired data to model the process of transformation from a clean audio input to a rendered audio output. Paired data includes unprocessed audio and the rendered audio with the same content setting, which is almost impossible to having both side of the paired data in a real world scenario.
Training in a supervised setting has already yielded  promising results for VA modeling and guitar amplifier modeling. Many commercial applications have adopted this approach, using mainly the minimization of the error-to-signal ratio (ESR) as the training objective \cite{769f627fa4fe49569bd207f6b1d32dc3, 8682805}. 
%While existing approaches 
%utilizing either a DSP-based model or neural network 
% particularly in guitar amplifier modeling tasks, are viable, 
However, the supervised approach does not scale well, for it requires paired data to model the transformation process from a clean audio input to a rendered audio output for a target tone. 
% Paired data comprises unprocessed audio and rendered audio with identical content settings, a scenario that is nearly impossible to achieve in real-world situations.
Each data pair has to be temporally aligned and be about the same content of guitar performance.
%paired data comprises clean audio and rendered audio with the same content settings. 
In many cases, 
%however, it can be difficult to have access to the clean version of a given rendered audio. 
%an aligned unprocessed audio corresponding to rendered audio is often missing and challenging to perfectly invert from the rendered audio.
%In real-world scenarios, 
however, the tone or timbre of a \emph{rendered} audio signal often lacks the \emph{unprocessed}, or direct-input (DI), audio counterpart, making supervised methods impractical. % due to their reliance on paired data during training. 
While it might be possible to create such data pairs by inverting the clean audio directly from a rendered audio by means of a guitar effect removal model, the development of such models is still an ongoing area of research \cite{imort2022distortion}.

In other audio synthesis tasks such as neural vocoding \cite{kumar2019melgan, kumar2024high} and voice conversion \cite{kaneko2019cyclegan, li2021starganv2}, 
many advanced generative adversarial network (GAN) \cite{goodfellow2020generative} models have been developed to generate realistic waveforms.
%the use of  has seen great success. 
For example, MelGAN \cite{kumar2019melgan} proposed a multi-scale discriminator (MSD) for distinguishing  between real audio and generated audio. HiFi-GAN \cite{kong2020hifi} proposed a multi-period discriminator (MPD) that collaborates with the MSD. 
%In addition to vocoder, GANs have been applied to voice conversion tasks \cite{kaneko2019cyclegan, li2021starganv2} without the need for parallel data. This approach treats voice conversion as an audio style transfer problem.
% that is, we can consider the voice conversion task as a audio style transfer problem.
Compared to models that minimize directly the reconstruction loss, GAN models employ such discriminators to learn customized loss functions in a data-driven fashion, usually leading to models that generate audio with finer details and better perceptual quality empirically.

%From a machine learning point of view, 
We conjecture that a GAN-based approach can similarly offer two advantages for guitar amplifier modeling.
% Although training in a supervised setting has already yielded very promising results. As demonstrated by \cite{769f627fa4fe49569bd207f6b1d32dc3, 8682805}, which were trained with the error-to-signal ratio (ESR) and are now widely utilized in various commercial products. We still believe that a GAN-based approach can offer two advantages in the amplifier modeling task.
\begin{description}
   
   \item[Adversarial losses] 
   % In comparison to the supervised loss (i.e., ESR), which often aims to achieve a mean value for fitting the distribution of the target, adversarial losses can directly minimize the Kullback–Leibler divergence between distribution generated audio and target audio.
   % Unlike supervised losses such as ESR, which often aims to fit the distribution of the target by updating the model according to the supervised loss, adversarial losses can directly minimize the Kullback–Leibler divergence between the distribution of generated audio and target audio through a minimax game.
    Adversarial losses offer a way to learn complicated, high-dimensional probability distributions from diverse and high-quality training %real 
    data samples without explicitly modeling the underlying probability density function. 
    %By training the discriminators with 
    Specifically, GANs implicitly learn the data distribution using a self-learned  loss function that is dynamically-adjusted as the training process unfolds.
    %that usually aligns better with human perception.
    \item[Unsupervised training] 
   % We can utilize any available clean data as input to the generator, with the target being the amplifier-rendered data, for the training process.
   We can use any available unpaired clean data as input to the generator during the training process, with the target being the designated amplifier-rendered data, thus potentially improving the generalizability of the model while reducing the burden of collecting paired data.
\end{description}

% Similar to guitar amp modeling task, a related research problem is vocoder task, which is also aim to generate the high fidelity raw audio waveform. [MelGAN] introduce a multi-scale discriminator which operate on different audio scales(i.e., downsampled audio). This multi scale setting enable each discriminator extract feature from different frequency range, since audio is losing high frequency information after the downsampling process according to Nyquiest theorem. These feature are used to distinguish between the label raw audio and generated raw audio. Even though the input of vocoder is often a mel-spectrum, it can still be consider as an audio-to-audio problem. 

% The progress of improving vocoder often comes up with a new architecture of discriminators design. [Hifi-GAn] added an additional multi-period-discriminator capture implicit feature between along several sample point in a set of distance. In [Avocode], they indicated that GAN-based model is suffer from artifact issues (e.g., alias). They proposed a collaborative multi-band discriminator (CoMBD) and a sub-band discriminator that helps model instead only learned a part of frequency but both low ad high frequency components.
% As training with a supervised setting has already lead to a very promising result, \cite{769f627fa4fe49569bd207f6b1d32dc3, 8682805} that are trained with error-to-signal ratio (ESR) were widely used in several commercial products in these days. We still assume that a GAN-based approach can possess two advantages in amplifier modeling task.

To the best of our knowledge, the work of Wright \emph{et al.} \cite{wright2023adversarial} represents the first and the only existing work that adopts GANs for guitar amplifier modeling.
%were the first to introduce unsupervised GANs to a guitar amplifier modeling task. 
Viewing  amplifier modeling as a style transfer problem, they showed that a GAN-based model using the MSD proposed in MelGAN~\cite{kumar2019melgan} as the discriminator can learn the  amplifier modeling process without using reconstruction loss functions such as the ESR.
%in an unsupervised setting. 
Moreover, they conducted experiments involving mismatched guitar timbre conversion between two timbres produced from distinct guitars. These experiments demonstrate the potential of adapting the unsupervised approach for guitar amplifier modeling.
%\ericrevised{[Write a bit more their exp setting and their accomplishments/findings.]}

Being inspired by the work of Wright \emph{et al.} \cite{wright2023adversarial}, we set forth to further extend this GAN-based approach by presenting the following two extensions.
%While demonstrating encouraging initial results, we believe that their work can be extended in many directions.
First, while they position their work in the context of audio \emph{style transfer} \yuhuarevised{by employ MelGAN\cite{kumar2019melgan} and several choices of spectral discriminator,} 
% we note that a variety of more advanced discriminators have  been proposed in research on \emph{neural vocoders}, 
\yuhuarevised{we further highlight the potential of integrating more advanced discriminators as proposed in neural vocoder research.}
% which aim to recover audio waveforms from given Mel-spectrograms. 
For example, it is well known that HiFi-GAN \cite{kong2020hifi} empirically generates audio waveforms with higher quality than MelGAN~\cite{kumar2019melgan}.
Research on neural vocoders is relevant, because both vocoders and guitar amplifier modeling aim to produce high-quality audio waveforms given some input conditions. 
Consequently, our first extension replaces the MSD discriminator  used in \cite{wright2023adversarial} by a combination of MSD and MPD discriminators, to study whether advanced discriminators can similarly contribute to better result for guitar amplifier modeling as the case seen in neural vocoding.

%In our study, we delve into exploring potential enhancements based on the framework proposed by Wright \emph{et al.} \cite{wright2023adversarial}.
%We replaced the MelGAN discriminator as described in \cite{wright2023adversarial} with a HiFi-GAN \cite{kong2020hifi} discriminator and observe that modifications are necessary to adapt the discriminator to the guitar amplifier modeling task and dataset. 

Our second contribution investigates more deeply the benefits of a GAN-based model in utilizing unpaired data. Specifically, we note that during the training process, Wright \emph{et al.} \cite{wright2023adversarial} only used the unprocessed audio that \emph{do} have the corresponding rendered audio of the target tone as the input to the generator.
%\yuhuarevised{Moreover, only a single MSD discriminator was applied in their GANs, which consecutively discriminate audio samples at different levels (i.e., sample rates).}
% \ericrevised{Moreover, ...}
However, as the GAN training does not require paired data, it is actually possible to utilize unprocessed audio that \emph{do not} have the rendered audio counterpart of the target tone as the generator's input. 
We study such a case in our work, using input audio signals that do not align with the target output audio signals in training our  model.
%As such, we conduct an experiment leveraging clean tone raw audio from multiple datasets to determine whether the performance of the unsupervised setting could benefit from this approach.

% \yuhuarevised{core contirubtion Mentioned 2-page demo paper we further make an extension....}
% \yuhuarevised{Demo page}

%The contribution of the paper is two-fold. Firstly, it enhances the performance of unsupervised amplifier modeling \yuhuarevised{by adapting a MSD and MPD.}. Secondly, by integrating clean audio from other dataset, further improvements can be achieved, particularly for very high-gain target tones. This study also serves as an extension of our previous demo paper \cite{chen2023neural}. 
We conduct experiments on two public-domain guitar datasets, the \emph{EGDB} dataset~\cite{chen2022towards} that have both low-gain and high-gain tones, and the \emph{EGFxset} dataset~\cite{pedroza2022egfxset} for an extremely high-gain tone.
%Finally, instead of testing on a single dataset, we consider two datasets in this work, one featuring low-gain guitar amplifiers and the other high-gain amplifiers. This helps investigate whether the GAN approach benefits the modeling of low-gain or high-gain amplifiers more.
Experimental results show that the proposed extensions contribute positively to the modeling result, especially for the extremely high-gain case.
We provide audio samples online.\footnote{\url{https://ampDaFX24.notionlinker.com}}

% The paper is structured as follows: Section 2 reviews neural amplifier modeling methods and GANs. Section 3 presents our proposed method, and the details of experiments aare explained in Section 4. Objective evaluation results are reported in Section 6. Finally, Section 7 concludes with a discussion of future work in neural amplifier modeling.

The paper is structured as follows: Section 2 reviews neural amplifier modeling methods and GANs. Section 3 presents our proposed method.
Section 4 describes the dataset and experimental setup; Section 5 reports the objective evaluation results. Section 6 discusses the results further. Finally, Section 7 concludes the paper with some ideas of future work.

\section{Related Works}

% In this paper, we formed the guitar amplifier modeling into a high-fidelity audio waveform generation task.  Given an input audio of T samples, $ x \in \mathbb{R}^{1*T}$ and a neural networks generator $ g $. We expect this generator $ g $ can carry out the amplifier modeling process. Since we are using neural network as our generator, we consider out method as a "black-box" approach.  

\subsection{Neural Amplifier modeling}

% Thanks to the development of deep learning, the use of neural networks on neural amplifier has been applied in several works on amplifier modeling\cite{769f627fa4fe49569bd207f6b1d32dc3, 8682805, wright2022grey, eichas2018jaes,6567472, Zhang2018AVG}. The neural network approach share some similarities to traditional black-box method. For example, a convolutional layer can be considered as a Wiener model followed by a static non-linearity. However, these techniques was original designed for speech-related task which are often in a 16kHz setting.  Since a overdrive or distortion sound characteristic sound often occurred on the higher frequency part. To achieve high fidelity audio, we need to deal with a relative higher sample rates compares to speech (e.g., 44.1 khz or 48khz) in order to retain high frequency information. That means the complexity and size of the model will increased which makes this task harder. 
Thanks to advancements in deep learning, neural networks have been utilized in several studies on amplifier modeling \cite{769f627fa4fe49569bd207f6b1d32dc3, 8682805, wright2022grey, eichas2018jaes,6567472, Zhang2018AVG}. The neural network approach shares similarities with traditional black-box methods.  For example, a convolutional layer can be conceptualized as a Wiener model \cite{HMModel}. Existing neural network models for amplifier modeling are usually adapted from neural network models that are initially proposed for speech-related tasks.
While speech signals commonly operate at a 16\,kHz sampling rate, overdrive or distortion sound characteristics frequently manifest in the higher frequency range, requiring sampling rates of 44.1\,kHz or 48\,kHz.
As such, the adaptations may result in increased complexity and model size, which can be unfavorable given the requirements on real-time efficiency and low latency  of VA modeling.
%some adaptation is usually needed for amplifier modeling.
%achieving high-fidelity audio entails working with relatively higher sample rates compared to speech (e.g., ) to preserve high-frequency information. This requirement results in an increased complexity and size of the model, thereby making the task more challenging.

\subsection{Generative Adversarial Networks}
A GAN~\cite{goodfellow2020generative} is a generative model contains two components: a generator $G$ and a discriminator $D$. The discriminator $D$ is essentially a classifier and it aims to output a value close to 1 for samples from ``real'' data distribution $x \sim p_{d}$,  and a value close to 0 for \textit{``}fake\textit{''} samples $G({z})$ generated by the generator $G$, whose input $z$ is sampled from a prior  distribution $p_{z}$. On the other hand, the generator seeks to deceive the discriminator by generating samples that are indistinguishable from real ones. 
% The formal loss equations are defined as follow: 
The two-player minimax game with the value function $V(G, D)$ is defined as follows, updating $G$ and $D$ iteratively as the training unfolds,

% \min _G \max _D V(D, G)=\mathbb{E}_{\boldsymbol{x} \sim p_{\text {data }}(\boldsymbol{x})}[\log D(\boldsymbol{x})]+\mathbb{E}_{\boldsymbol{z} \sim p_{\boldsymbol{z}}(\boldsymbol{z})}[\log (1-D(G(\boldsymbol{z})))]

\begin{equation}
    \begin{aligned}
    \min _G \max _D V(D, G)=\mathbb{E}_{\boldsymbol{x} \sim p_{\text {d }}(\boldsymbol{x})}[\log D(\boldsymbol{x})]+ \\
    \mathbb{E}_{\boldsymbol{z} \sim p_{\boldsymbol{z}}(\boldsymbol{z})}[\log (1-D(G(\boldsymbol{z})))]
    \end{aligned}
	\label{generator_loss}
\end{equation}

For VA modeling, it is the generator $G$ that performs the clean-to-rendered transformation during both the training and inference stages. The discriminator $D$ only functions during the training stage, guiding how the generator is optimized. Therefore, it is possible to use a computationally heavy discriminator to train a light generator, for better run-time efficiency of the generator.

% \begin{equation}
% 	\min _G-\mathbf{E}_{{z} \sim p_{{z}}}[\log D(G({z}))]
% 	\label{generator_loss}
% 	\end{equation}
% 	\begin{equation}
%     \max _D \mathbf{E}_{{x} \sim p_d}[\log D({x})]+{E}_{{z} \sim p_{{z}}}[\log (1-D(G({z})))]
%     \label{discriminator_loss}
% \end{equation}

\subsection{Backbone Model for Generator}
% The backbone model of the generator is a wavenet model, which is widely apply in many audio related task(e.g., vocoder, VA modeling). 

Existing approaches to neural VA modeling can be categorized into two main types: convolutional (CNN) and recurrent neural networks (RNN).
From a digital signal processing  viewpoint, CNN networks can be viewed as finite impulse response (FIR) filters.
CNN-based models have demonstrated superior performance in modeling various devices. For example, Wright \emph{et al.}~\cite{wright2020real} applied a WaveNet model~\cite{oord2016wavenet}  to model the Blackstar HT-5 Metal and the Mesa Boogie 5:50 Plus amplifiers.
Damsk{\"a}gg \emph{et al.}~\cite{damskagg2019deep} utilized a WaveNet model with conditioning control on the gain parameter to emulate the Fender Bassman 56F-A vacuum-tube amplifier.
In addition to amplifier modeling tasks, Steinmetz \emph{et al.}~\cite{steinmetz2021steerable} trained a conditional temporal convolutional network on compressor, analog delay, guitar amplifier, and reverberation effects. 

On the other hand, RNN-based approaches often rely on long-short term memory (LSTM) or gated recurrent units (GRU). For instance, Wright \emph{et al.}~\cite{769f627fa4fe49569bd207f6b1d32dc3} showed promising results using a recurrent-based model to model a high-gain channel Blackstar HT-1 vacuum tube amplifier and an Electro-Harmonix Big Muff Pi distortion/fuzz pedal.  Juvela \emph{et al.}~\cite{juvela2023end} extended their work further by concatenating control parameters with a range of [0,1] as additional input channels to their LSTM network.

% \cite{steinmetz2021steerable} 

% The architecture of the convolution design enable wavenet model to capture the dependency of time-related feature which is crucial in raw waveform audio domain. 

% [firstwavenetpedal], they applied a non-casual wavenet to a guitar amplifier modeling. Is is widely apply in many audio related task(e.g., vocoder, VA modeling). [alec icassp19] use a wavenet model consists of two stacks of nine dilated convolution for their generator under GAN framework.  

% RNN PART

\subsection{Discriminators for GANs training}
% To apply adversarial losses under GANs framework, we need a discriminator to distinguish between the real data and generated output. The GANs framework was applied priory on different speech task such as vocoder and voice conversion. In \cite{kumar2019melgan}, they built a multi-scale discriminator (MSD) which operate on different audio scales (i.e. sample rates). Each scale's audio has it's own 1-D convolution based module to get the output.The output of each scale's output were used to calculate on the adversarial losses to train the discriminator. In \cite{kong2020hifi}, they added an multi-period discriminator (MPD) with MSD to capture both regular and prime number distance between each sample points. For spectral-based discriminators, \cite{yamamoto2020parallel} introduced a multi-resolution STFT auxiliary loss. They calculated and sum the n short-time Fourier transform (STFT) losses with different parameters (i.e., FFT size, window size, and frame shift). These techniques make a generator not only focus on generating waveform itself but also need to generate reasonable results under spectral-domain.

To apply adversarial losses within the GAN framework, a discriminator is needed to distinguish between real data and generated output. 
\woosungvised{
% GANs have been applied to various audio tasks such as vocoder \cite{kumar2019melgan, kong2020hifi}, voice conversion \cite{kaneko2019cyclegan, li2021starganv2}, and neural codec \cite{defossez2022highfi}.
Several discriminators have been proposed for audio generative tasks such as neural vocoder \cite{kumar2019melgan, kong2020hifi}, voice conversion \cite{kaneko2019cyclegan, li2021starganv2}, and neural codec \cite{defossez2022highfi}.
We categorize these discriminators into two types: spectral-based and waveform-based discriminators.}

For spectral-based discriminators,  Défossez \emph{et al.} \cite{defossez2022highfi} proposed a a multi-scale STFT-based discriminator. They computed and summed the short-time Fourier transform (STFT) losses with different parameters (i.e., FFT size, window size, and hop length). These techniques compel the generator to not only focus on generating the waveform itself but also to generate reasonable results in the spectral domain.
On the other hand in MelGAN \cite{kumar2019melgan}, a multi-scale discriminator (MSD) was introduced, operating on different audio scales (i.e., sample rates) \woosungvised{in waveform domain}. Each scale's audio is processed by a 1-D convolution-based module to obtain an output. The outputs from each scale are then used to calculate adversarial losses for training the discriminator. 
% In Hifi-GAN \cite{kong2020hifi}, a multi-period discriminator (MPD) was added to the MSD to capture both regular and prime number distances between sample points. 
\woosungvised{
In Hifi-GAN \cite{kong2020hifi}, a multi-period discriminator (MPD) was proposed to capture both regular and prime number distances between sample points, resulting in improved speech synthesis quality.} 
While the previous GAN-based VA modeling model  \cite{wright2023adversarial} uses MSD alone, we use MSD \woosungvised{as well as} MPD in our work.

% For spectral-based discriminators, \cite{yamamoto2020parallel} introduced a multi-resolution STFT auxiliary loss. They computed and summed the short-time Fourier transform (STFT) losses with different parameters (i.e., FFT size, window size, and frame shift). These techniques compel the generator to not only focus on generating the waveform itself but also to generate reasonable results in the spectral domain.

\begin{table}[]
\begin{adjustbox}{width=\columnwidth}
\begin{tabular}{c|cccc}
                           & GuitarSet & EGDB & GUITAR-FX-DIST & EGFxSet  \\ \hline
Clean    & ~~3h        & ~~2h  & ~~0.57h         & ~~$\sim$1h \\
Rendered & N/A       & 10h & $\sim$111h    & 11.5h  
\end{tabular}
\end{adjustbox}
\caption{The total duration (in hours, or `h') of the clean, unprocessed audio and the rendered audio (with effects applied) of four existing public paired datasets, GuitarSet~\cite{xi2018guitarset}, EGDB~\cite{chen2022towards}, GUITAR-FX-DIST~\cite{comunita2020guitar}, and EGFxSet~\cite{pedroza2022egfxset}.}
%Clean denotes the duration of clean audio in each dataset. Rendered denotes the total duration of all effects in each dataset.
\label{tab:dataset_comparison}
\end{table}

\subsection{Clean Audio from Existing Datasets} 

We refer to a dataset as a \emph{paired dataset} when it contains the clean audio signal counterpart for each amplifier- or pedal-rendered audio signal. Referring to existing public-domain paired datasets, we show a comparison of clean audio and rendered audio duration in Table \ref{tab:dataset_comparison}. As the table shows, the duration of amplifier or pedal rendered data in each dataset is much longer than the aligned clean audio in overall duration. 
We see that clean audio is relatively scarcer than rendered audio.
%For instance, the dataset presented in EGDB \cite{chen2022towards}, originally designed for automatic transcription tasks and using the same device to capture clean audio input, has a duration of 2 hours due to manual annotation processes.

We categorize the clean audio into two types during training. First is \emph{target-aligned} clean audio, where a target audio can always find an aligned clean audio with the same musical content. % but in a clean setting.
Second is \emph{target-unaligned} clean audio, where the musical content in this type of clean audio does not exist in target amplifier-rendered audio. 
\yuhuarevised{Please note that the objective evaluation metrics such as 
ESR and Mel-spectrum loss (cf. Section \ref{metrics})
%based on the distance of features 
still requires data from a target-aligned setting.}

As mentioned in Section \ref{sec:intro}, \yuhuarevised{even though the clean data and rendered data from the \emph{same} dataset is usually fully aligned (i.e., they are target-aligned),} 
we can take advantage of the GAN-based approach and further use clean data and rendered data from \emph{different} datasets  and employ such target-unaligned data in our unsupervised training. 
%even in the absence of target pairs. 
\ericrevised{The prior work of Wright \emph{et al.} \cite{wright2023adversarial} did not exploit such a potential, 
as they used clean audio from \cite{kehling2014automatic} and created rendered target audio from three different plugins, essentially creating target-aligned data. 
Unlike their work, we study the use of target-unaligned data in our experiments.}

% The model
% has two main components, a stack of dilated causal convolutional
% layers, and a linear post-processor. The post-processor is a fully
% connected 

\section{Methods}
\begin{figure*}[ht!]
\center
\subfigure{\includegraphics[width=\textwidth]{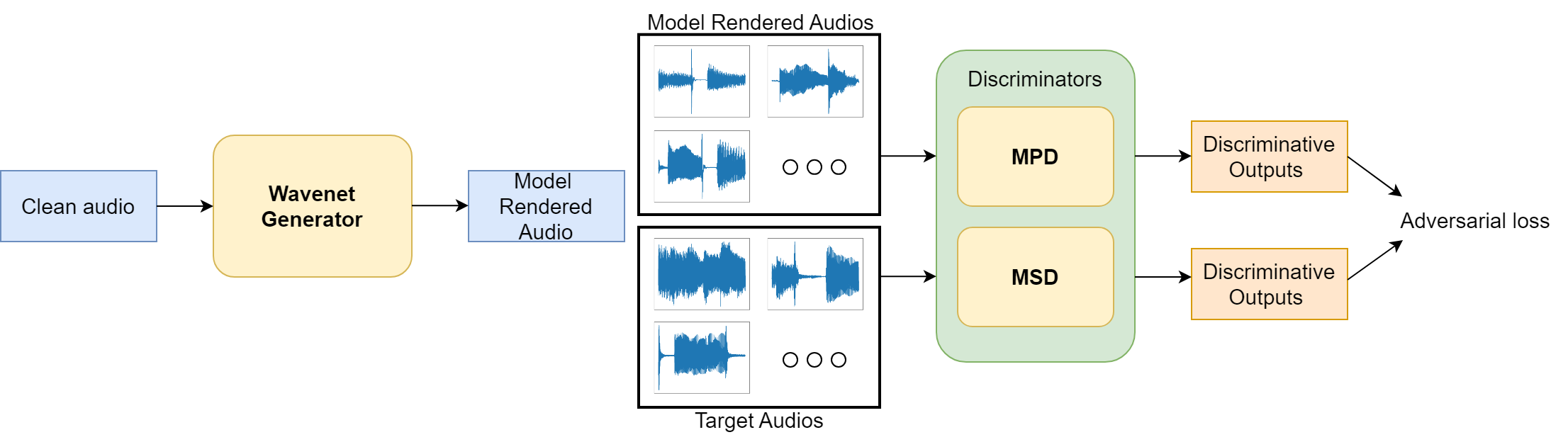}} 
\caption{\label{model_arch}{Diagram of the proposed GAN-based model for VA modeling, using clean audio that may not be matched and aligned with the target audio segment, and using two types of discriminators: MSD and MPD~\cite{kong2020hifi}.}}
\end{figure*}

% In this work, We employ a GAN-based framework. A generator and discriminator is trained are trained adversarially. 
We consider guitar amplifier modeling as a generative task that aims to generate high-fidelity audio waveforms.  Given an input audio of $T$ samples, $\mathbf{x} \in \mathbb{R}^{1\times T}$, 
we adopt the ``black-box'' approach and train  a neural network-based generator $G$ that carries out the amplifier modeling process and generates $\hat{\mathbf{x}}=G(\mathbf{x})$. 
%Since we use neural network as our generator, we consider out method as a ``black-box'' approach. 
We illustrate our training framework in Figure~\ref{model_arch}.

\subsection{Generator}
\label{generator_method}
We employ the same causal feed-forward WaveNet model architecture as Wright \emph{et al.} \cite{wright2023adversarial} for our generator.  It consists of two stacks of nine dilated convolution layers. The dilation is one at the first stack and is increased by a factor of two after each stack to get a larger receptive field. We set a kernel size as 3 to get a growth receptive field from small area to larger area. Each convolution layer is equipped with a weight normalization. We use the same gated activation function as the original WaveNet model~\cite{oord2016wavenet}. 

% Following the generator setting in \cite{wright2023adversarial}, we employed the same architecture using a causal feed-forward WaveNet model as our generator. It consists of two stacks of nine dilated convolutions. The dilation is set to one in the first stack and is increased by a factor of two after each stack to achieve a larger receptive field. We set the kernel size to 3 to ensure a gradual increase in the receptive field from a small area to a larger one. Each convolution layer is equipped with a weight normalization layer. Additionally, we use the same gated activation function as the original WaveNet \cite{van2016wavenet}. Please refer to the figure below for visualization:

\begin{table}[] 
\begin{adjustbox}{width=\columnwidth}
\begin{tabular}{c|ccccc}
Model  & channels     & kernel sizes & stride & groups & padding \\ \hline
conv1d & ~~~(1, 128)     & 15           & 1      & 1      & 0       \\
conv1d & (128, 128)    & 41           & 2      & 4      & 20      \\
conv1d & (128, 256)    & 41           & 2      & 16     & 20      \\
conv1d & (256, 512)   & 41           & 4      & 16     & 20      \\
conv1d & (512, 1024)  & 41           & 4      & 16     & 20      \\
conv1d & (1024, 1024) & 41           & 1      & 16     & 20      \\
conv1d & (1024, 1024) & 5            & 1      & 2      & 0      
\end{tabular}
\end{adjustbox}
\caption{Parameter settings of the convolutional layers of the implemented MSD sub-discriminators.}
\label{MSD_table}
\end{table}

\begin{table}[] 
\begin{adjustbox}{width=\columnwidth}

\begin{tabular}{c|ccccc}
Model  & channels     & kernel sizes & stride & groups & padding \\ \hline
conv2d & (1, 32)     & (5, 1)           & (3,1)      & 1      & 2      \\
conv2d & (32,128)    & (5, 1)           & (3,1)      & 1      & 2      \\
conv2d & (128, 512)    & (5, 1)           & (3,1)      & 1     & 2      \\
conv2d & (512, 1024)   & (5, 1)           & (3,1)      & 1     & 2      \\
conv2d & (1024, 1024)  & (5, 1)           & (1,1)      & 1     & 2      \\
conv2d & (1024, 1)~~~~~ & (1,1)           & (3,1)          & 1     & 2      \\
\end{tabular}
\end{adjustbox}
\caption{Parameter settings of the convolutional layers of the implemented MPD sub-discriminator.}
\label{MPD_table}
\end{table}

\subsection{Discriminator}
\label{discriminator_method}
% To capture the information and the dependency is crucial for distinguish between real and generated outputs. Therefore we applied two modules: MSD and MPD from [Hifi GAN] discriminators. \n
%To capture the information and the dependency is crucial distinguishing between real and generated outputs. We apply two modules: MSD and MPD \cite{kong2020hifi}.
Our discriminator consists of both MSD- and MPD-based ones.
The MSD consists of three sub-discriminators originally \cite{kong2020hifi}. 
\woosungvised{However, we remove the last sub-discriminator that processes the audio after two downsampling layers (i.e., the one that operates at the lowest temporal resolution), as this gives better results empirically in our pilot study.}
%to a number of two sub-discriminators in MSD. 
The input flow for MSD is therefore: raw audio, the first sub-discriminator, $\times 4$ average-pooled audio, and finally the second sub-discriminator. We set the parameters as shown in Table~\ref{MSD_table}. Following the setting of Hifi-GAN \cite{kong2020hifi}, spectral normalization is applied for the first sub-discriminator, while weight normalization is applied for the second one.

% \n The MPD is consists of mixture sub-discriminators. On the other side of MSD, it only accepts equally distance sample points of the input audios. With audio length $T$ and distance $P$, the shape of input for each sub-discriminator will reshape from the audio length $T$ to $(T/P, P)$ (i.e., from 1D to 2D). 
% The distance is given by a set of parameters, we follow \cite{kong2020hifi} for their setting of the parameters: [2, 3, 5, 7, 11]. Each sub-discriminator is a stack of convolutional layer, weight normalization is applied for every sub-discriminator. We can get a set of discriminative output of each distance sub-dicriminator for a different perspective result from different distance setting. 

% \begin{figure*}[ht!]
% \center
% \subfigure{\includegraphics[width=1\columnwidth]{DAFx24_Templates_LaTeX/Image/MPD.drawio.png}} 
% \caption{\label{model_arch}{Our GANs framework }}
% \end{figure*}

% The MPD consists of mixture sub-discriminators.
The MPD comprises a collection of mixture sub-discriminators.
Unlike MSD, it only accepts equally-spaced sample points of the input audios. With audio length $T$ and period $P$, input for each sub-discriminator will be reshaped from the audio length $T$ to $(T/P, P)$ (i.e., from 1D to 2D).
% Multiple possible periods $P$ are considered, including $P=[2, 3, 5, 7, 11]$, again following the setting of Hifi-GAN \cite{kong2020hifi}. 
\woosungvised{Following the setting of Hifi-GAN \cite{kong2020hifi}, 
 we employ multiple sub-discriminators, each operating with a period $p$ in $P=[2, 3, 5, 7, 11]$.}
We set the parameters of the convolutional layers as shown in Table~\ref{MPD_table}.
Each sub-discriminator is a stack of convolutional layers, with weight normalization applied for every convolutional layers. This setup allows us to obtain a set of discriminative outputs for each period sub-discriminator, providing different perspectives based on different period settings.

There are also other voice synthesis works under GANs framework collaborating their discriminators with spectral-based discriminator \cite{jang2021univnet, yang2020vocgan}. Either a hierarchical \cite{yang2020vocgan} or multi-resolution \cite{jang2021univnet} approach aims to capture the information from different perspectives of sample rates or window sizes of short-time Fourier transform. 
% \cite{wright2023adversarial} also apply a spectral-based discriminator 
% The reason we choose to follow several settings in \cite{kumar2024high} is not only because of it's success in speech synthesis. At our preliminary experiment, we found that applying MPD as a module of discriminator can help the modeling of higher frequency information in distortion or overdrive audio. Furthermore, the combination of MSD$+$MPD was not only applied on vocoder task but also on neural codec task\cite{kumar2024high}. To be summarize, the main goal of these audio-related task are all aim to generate a high fidelity sound and this discriminator combination can improved the audio fidelity under GAN training.
% The reason we chose to follow several settings in \cite{kong2020hifi} is not only because of its success in speech synthesis.
We opt to follow several settings from \cite{kong2020hifi} not solely due to its success in speech synthesis.
In our preliminary experiments, we found that applying MPD as a module of the discriminator can help model high frequency information in distortion or overdrive audio. Furthermore, the combination of MSD$+$MPD has been not only applied in neural vocoder tasks but also in neural codec tasks \cite{kumar2024high}. In summary, the main goal of these audio-related tasks is to generate high-fidelity sound, and utilizing multi-type discriminators can improve audio fidelity under GAN training.

\subsection{GAN Loss}

% In \cite{kong2020hifi}, they apply LS-GAN \cite{mao2017least} as their generator and discriminator training loss functions for non-vanishing gradient flows. 
We choose to apply a Hinge GAN loss function \cite{lim2017geometric} in our GAN training, due to its promising result in prior work \cite{kumar2024high,wright2023adversarial}. The loss equation is defined as follows:

\begin{equation}
\mathcal{L}(D;G)=\mathbb{E}_y[\max (0,1-D(y))]+ 
\mathbb{E}_x[\max (0,1+D(G(x)))]
\label{eq1}
\end{equation}

\begin{equation}
\mathcal{L}(G;D)=\mathbb{E}_x[-D(G(x))]
\label{eq2}
\end{equation}

% During the training, the discriminator is trained to classify label data $y$ to 1, and the samples generated from Generator $G(x)$ to 0. The generator is trained to deceive the discriminator to recognize the $G(x)$ as a real label data, that is close to 1.

\woosungvised{
During training, the discriminator is trained to classify labeled data $y$ as 1, and the samples generated from Generator $G(x)$ as 0. The generator is trained to deceive the discriminator into recognizing $G(x)$ as real  data, aiming for a classification close to 1.
}
% In \cite{kumar2024high,larsen2016autoencoding, kumar2019melgan}, there are other auxiliary loss such as \emph{mel-spectrogram loss} and \emph{feature matching loss}. These losses require a requirement of paired data setting during the training process. A \emph{mel-spectrogram loss} calculate the L1 distance between mel-spectrogram of the generated audio and a that of a label data. A \emph{feature matching loss} calculate the L1 distance in features of the intermediate discriminators between the generated audio and a label data.
\woosungvised{
Other auxiliary losses such as \emph{mel-spectrogram loss} and \emph{feature matching loss} were used in \cite{kumar2024high, larsen2016autoencoding, kumar2019melgan}.
% These losses require a paired data setting during the training process.
The mel-spectrogram loss measures the L1 distance between the generated audio's mel-spectrogram and that of labeled data. 
The feature matching loss computes the L1 distance in intermediate features from the discriminators between the generated audio and a labeled data.
% Although these loss function can improve the training efficiency and stability of the generator also the quality of the generated audio, the requirement of paired data setting is not suitable in our cases. At the end, our model only use an adversarial loss during the training process for both generator and discriminator. 
Although these loss functions can improve the training efficiency, stability of the generator, and the quality of the generated audio, 
they require a paired data setting during the training process.
Given an unpaired data setting, our model only utilizes an adversarial loss during the training process for both the generator and discriminator.
}

\section{Experimental Setup}
\subsection{Dataset}
\label{dataset}
% We choose two electric guitar dataset for our experiments: \textbf{EGDB}\cite{chen2022towards} and \textbf{EGFxset}\cite{pedroza2022egfxset}. For EGDB, the duration of a single tone is approximately 2 hours. We choose a subset consists of \textbf{Marshall JCM2000}, \textbf{Fender Twin Reverb}, \textbf{Mesa Boogie Mark}. For EGFxset, the duration of a single tone is approximately 57 min. We choose \textbf{BD-2} for the dataset in our experiments.
We select two electric guitar datasets for our experiments: EGDB \cite{chen2022towards} and EGFxset \cite{pedroza2022egfxset}.
For EGDB, the duration of a single tone is approximately 2 hours. We choose a subset consisting of \textbf{Marshall JCM2000}, \textbf{Fender Twin Reverb}, and \textbf{Mesa Boogie Mark}.
For EGFxset, we select the \textbf{BD-2} dataset as the target tone for our experiments. As the gain value has been set to its maximum value for BD-2, this dataset contains highly distorted sound.
As the BD-2 tone has fairly high gain, we may consider the three tones from EGDB as relatively low-gain tone compared to BD-2.  
% \yuhuarevised{
% We note that, while the music content in the EGDB consists of musical phrases or licks, the music content in EGFxset comprises recordings of individual notes, each at a different pitch and from different pickups.
% }
\woosungvised{We note that EGDB comprises musical phrases or licks, while EGFxset contains recordings of individual notes, each at different pitches and from various pickups.}

% In our preliminary experiments, we found that the amplitude difference of two dataset is crucial and sensitive for a GAN-based approach. We normalized both datasets by pyloudnorm\cite{steinmetz2021pyloudnorm}. We normalized the peak of each audio to -1 dB, then normalized each audio to -12 dB LUFS. Without the normalization process, the training will be extremely unstable which cause a failure during the early stage of training process. 
In our preliminary experiments, we found great differences in amplitude between the two datasets, possibly because they were collected under different device settings (e.g., guitar or audio interface) and recording environments. We found that GAN-based models is highly sensitive to differences in amplitude. 
To address this, we normalize both datasets using pyloudnorm \cite{steinmetz2021pyloudnorm}.
Specifically, we normalize the peak of each audio to $-1$ dB, and then normalized each audio to $-12$ dB LUFS. Without such a normalization, the training would be extremely unstable, leading to failures during the early stages of the GAN training process.

% We divided the dataset into training, validation, and test sets using an 80/10/10 ratio. For training, the input of clean data and the output of target tone data are randomly arranged in each batch for an unsupervised setting. In order to evaluate the model performance, since two datasets are aligned between the clean data and rendered data, the validation and testing is under a paired setting to calculate among all metrics.
We divide the dataset into training, validation, and test sets using an 80/10/10 ratio. For training, the input clean data and the output target tone data are randomly arranged in each batch for an unsupervised setting.
To evaluate the model performance, as the clean data and rendered data are aligned between the two datasets, validation and testing are conducted under a paired setting to calculate all metrics.

% 2024/3/4 modification to here
\subsection{Metrics}
\label{metrics}
% Traditionally, amplifier modeling often trained and evaluated under a error-to-singal (ESR) metric. For $N$ sample points, a pre-emphasis filter was applied to the generated signal $\widehat{y}_{p}$ and a target signals $y_p$ before computing the ESR. The denominator of target signal itself is used to prevent the dominate by the high energy segments. 

We consider the following three metrics for objective evaluation. % in our performance study.

\textbf{Error-to-signal ratio} (ESR) is a metric commonly employed for training and evaluating an amplifier modeling model. For $N$ sample points, a pre-emphasis filter is applied to both the generated signal $\widehat{\mathbf{y}}_{p}$ and a target signals \(\mathbf{y}_p\) before computing the ESR. 
% \begin{equation}
$$
    {\mathrm{ESR}}=\frac{\sum_{n=0}^{N-1}\left|\mathbf{y}_p[n]-\widehat{\mathbf{y}}_p[n]\right|^2}{\sum_{n=0}^{N-1}\left|\mathbf{y}_p[n]\right|^2} \,.
    \label{esr_equation}
% \end{equation}
$$
The denominator of target signal itself is used to prevent the high energy segments from dominating the result. 
%Additionally, we examine our model using two other losses: 1. mel-spectrum loss 2. Fréchet Audio Distance(FAD). 

\textbf{Mel-spectrum loss} ($L1_{mel}$)
\label{mel_spec_loss}
measures the difference between the ground-truth and predicted audio in the spectral domain. Denoting $\phi(\cdot)$ as the function that converts an audio waveform into a Mel spectrogram, this loss can be calculated as follows,
$$L1_{{mel}}=\mathbb{E}_{(\mathbf{x})}\left[\|\phi(\mathbf{x})-\phi(\hat{\mathbf{x}})\|_1\right]\,. $$
%We calculate the L1 distance between the generated output and label data. 
% Since the effect characteristic of overdrive and distortion are more easily observed from the perspective of spectral domain, we conjecture that ESR is not sufficient for evaluating the performance of modeling overdrive and distortion, while the Mel-spectrum loss might be more appropriate.
\woosungvised{Since the characteristics of certain effects such as overdrive and distortion are more easily observed from the perspective of spectral domain, we conjecture that Mel-spectrum loss offers a more suitable measure than ESR for evaluating the performance of modeling overdrive and distortion.}

%Mel-spectrum loss can be expressed as follows, where $x$ and $ xˆ$ respectively denote a ground-truth and predicted audio, φ(·) denotes the fourier transform function to a mel spectrogram.

\textbf{Fr\'echet Audio Distance}  (FAD) \cite{kilgour2018fr} measures 
the Fr\'echet distance between the distribution of embedding from a set of reference audios and those from the generated audios.
It was first proposed to evaluate a music enhancement task and was found to correlate well with human perception. The metric has been later applied to music generation tasks  (e.g., \cite{yeh22ismir,agostinelli2023musiclm, copet2024simple}) to indicate if the generated audio is plausible. We accordingly adopt it here as well. 
To provide different insights than traditional alignment-based metrics, we report the FAD\footnote{To compute FAD, we use an open-source implementation:  \url{https://github.com/gudgud96/frechet-audio-distance}} score of all models with the VGGish model. Samples with a low FAD score are expected to be more plausible.
\ericrevised{We note that there might be better alternatives than the VGGish model for computing the FAD scores \cite{gui2024adapting}, but we leave that as a future work.}

\subsection{Implementation details}
% For the generator and discriminator detail, we have introduced in Section~\ref{generator_method} and Section~\ref{discriminator_method}. Please note that we keep all the generator model in a identical structure and setting including supervised method to have a fair comparison between each methods. 
For detailed information on the generator and discriminator, please refer to Sections \ref{generator_method} and \ref{discriminator_method}. It is important to note that we use the same structures and settings for all generator models, including the supervised method, for fair comparison.

% We split every audio to two-second segment during the training. Every model is trained with a single 3090 GPU. Supervised and melGAN discriminator model are trained with the original paper's setting.
% Two of our proposed discriminator: MSD and MSD$+$MPD are trained with Adamw optimizer with an initializing generator learning rate 5e-5 and discriminator learning 1e-5. We set the generator learning rate empirically with a higher number to prevent the imbalance training process between generator and discriminator, since our generator model sizes is much smaller than the both proposed discriminator setting.

During training, we split every audio into two-second segments. Each model is trained using a single RTX 3090 GPU. The supervised and MelGAN discriminator models are trained with the settings described in their original papers.

For our discriminators, MSD and MSD$+$MPD, we employ the AdamW \cite{loshchilov2017decoupled} optimizer with an initial generator learning rate of 5e--5 and discriminator learning rate of 1e--5\yuhuarevised{, along with a weight decay of 0.01.} We set the generator learning rate empirically to a higher value to prevent an imbalance in the training process between the generator and discriminator, as our generator model sizes are much smaller than those of the proposed discriminators.

\subsection{Evaluation settings}
We conduct two experimental scenarios to evaluate the introduced three metrics mentioned in Section \ref{metrics}.
First, we compare our method with the supervised approach of Damsk\"agg \emph{et al.}~\cite{8682805} and the unsupervised GAN-based approach proposed by Wright \emph{et al.} \cite{wright2023adversarial}.
Second, to validate the advantage of unsupervised learning, we combine the clean audio from EGDB and EGFxset as the input to our generator (marked as ``both'' in Table~\ref{experiment_result}), no matter whether the target tone is from EGDB or EGFxset.
The generator for all the aforementioned models used the same architecture; the difference in each model setting lies in the training method and the loss functions applied (e.g., supervised or adversarial).

% & $L1_{\text{mel}}\downarrow$

\begin{table}[h]
% \begin{adjustbox}{width=\columnwidth}
\resizebox{\columnwidth}{!}{
\centering
\begin{tabular}{l|ll|lll}
Target tone & Input & Model & $L1_{{mel}} \downarrow$ & ESR$\downarrow$ & FAD$\downarrow$ \\ \cline{1-6}
  & EGFxset & Supervised~\cite{8682805} & 4.041 & 0.106 & 5.256 \\ 
  BD-2 & EGFxset & MSD & 1.874 & 0.164 & 1.900 \\
  (EGFxset) & EGFxset & MSD$+$MPD & 1.535 & 0.052 & 0.983 \\
  & both & MSD$+$MPD & \textbf{1.156} & \textbf{0.022} & \textbf{0.550} \\ \midrule
  & EGDB & Supervised~\cite{8682805} & 2.342 & \textbf{0.019} & 1.657 \\
  Marshall & EGDB & MSD & 2.660 & 0.229 & 1.410 \\
  (EGDB) & EGDB & MSD$+$MPD & \textbf{2.315} & 0.028 & \textbf{0.994} \\
  & both & MSD$+$MPD & 2.458 & 0.029 & 1.054 \\ \cline{1-6}
  & EGDB & Supervised~\cite{8682805} & \textbf{1.953} & \textbf{0.014} & 1.126 \\
  FTwin & EGDB & MSD & 2.302 & 0.072 & 0.878 \\
  (EGDB) & EGDB & MSD$+$MPD & 1.960 & 0.021 & \textbf{0.346} \\
  & both & MSD$+$MPD & 2.267 & 0.020 & 0.434 \\ \cline{1-6}
  & EGDB & Supervised~\cite{8682805} & 1.705 & \textbf{0.012} & 1.923 \\
  Mesa & EGDB & MSD & 2.158 & 0.137 & 1.674 \\
  (EGDB) & EGDB & MSD$+$MPD & \textbf{1.633} & 0.014 & 1.748 \\
  & both & MSD$+$MPD & 1.694 & 0.041 & \textbf{1.400} \\
\end{tabular}
}
\caption{Objective evaluation result of the supervised models~\cite{8682805} and GAN-based models (i.e., MSD alone and ours) for different target tones (i.e., an extremely high-gain tone from EGFxset and three tones from EGDB), using source signals from different datasets as the model input (i.e., using target-aligned audio, or using ``both'' target-aligned audio and target-unaligned audio [i.e., EGDB+EFGxset]). All the three objective metrics are the lower the better; best results highlighted in bold.}
\label{experiment_result}
\end{table}

\begin{figure*}[ht!]
\center
\subfigure{\includegraphics[width=0.29\textwidth]{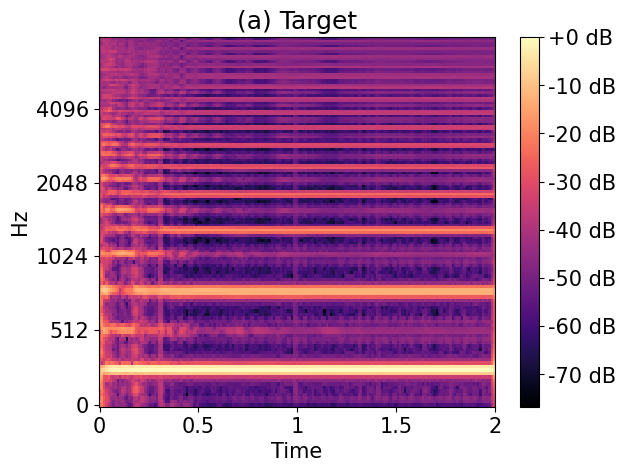}} 
\subfigure{\includegraphics[width=0.29\textwidth]{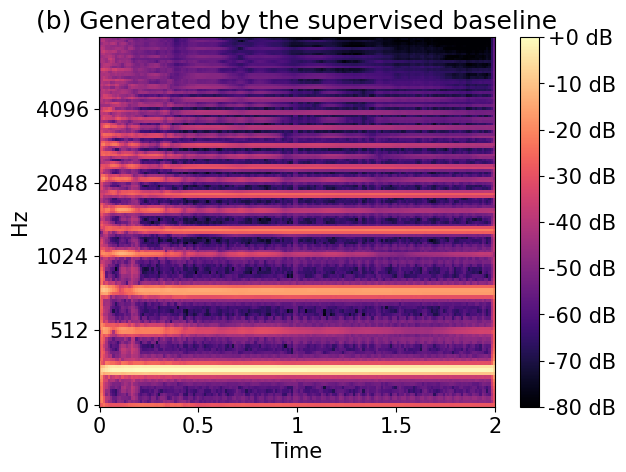}} 
\subfigure{\includegraphics[width=0.29\textwidth]{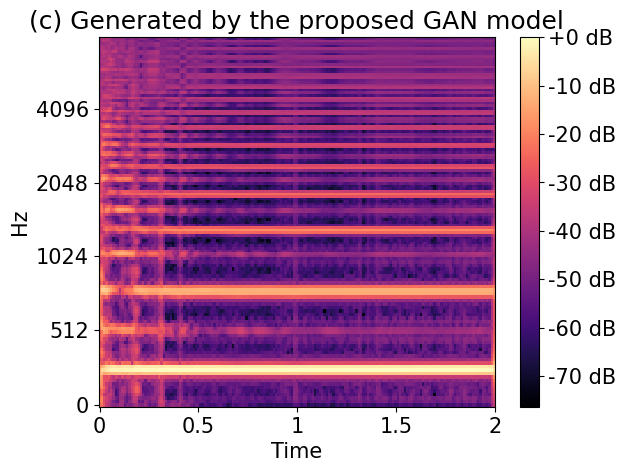}} 
\caption{\label{comparison_figures}{Mel-spectrogram of the target audio signal, along with the ones generated by the supervised baseline~\cite{8682805} and the proposed MSD$+$MPD GAN-based model given the corresponding clean audio signal. The target audio is sampled from the test set of EGFxset~\cite{pedroza2022egfxset}, with BD-2 being the target  tone. 
We see missing high-frequency harmonics from the top-right corner of the middle Mel-spectrogram, the one generated by the  supervised baseline.  }}
\end{figure*}

\begin{figure*}[ht!]
\center
\subfigure{\includegraphics[width=0.29\textwidth]{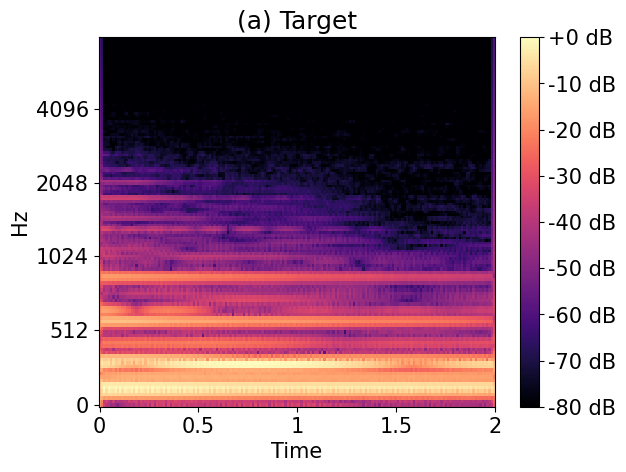}} 
\subfigure{\includegraphics[width=0.29\textwidth]{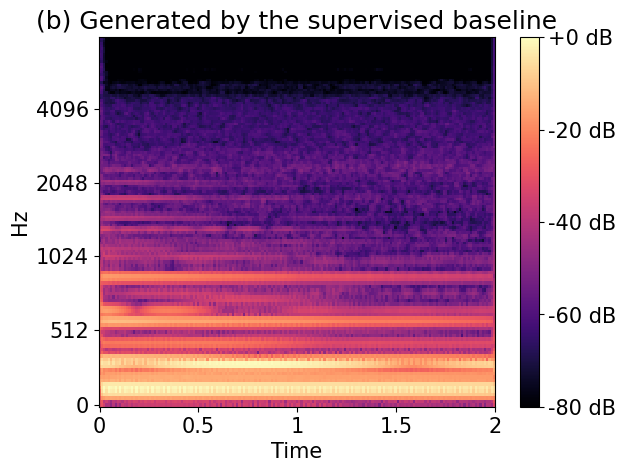}} 
\subfigure{\includegraphics[width=0.29\textwidth]{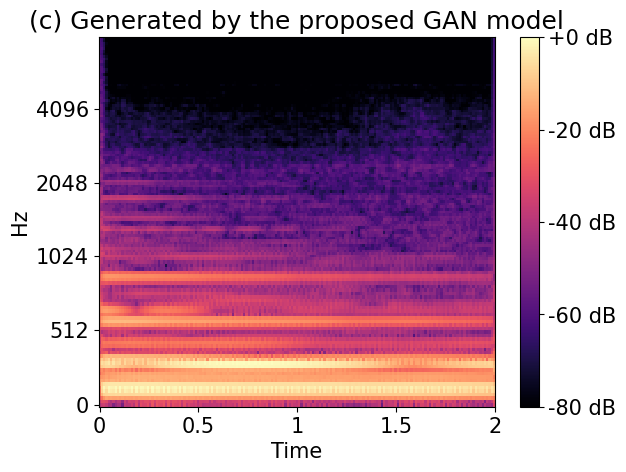}} 
\caption{\label{comparison_figures_2}{The mel-spectrogram between target audio , supervised approach and MSD$+$MPD sampled from the EGDB Fender test set. Both of supervised baseline and our MSD$+$MPD exhibit artifacts. These artifacts manifest as the generation of non-existent high-frequency information in the target mel-spectrogram. }}
\end{figure*}

\section{Experimental Result}
\subsection{Comparison with Baseline Methods}
Table \ref{experiment_result} shows the results of modeling different target tones.
We consider firstly the case when the input audio are from the same dataset as the target audio, namely using EGFxset input when the target tone is BD-2, and using EGDB input when the target tone is the other three. We will consider the result for the case when using input from ``both'' EGFxset and EGDB in the next subsection.

\yuhuarevised{As we address a challenging case with BD-2 as the target tone, which is characterized by a highly distorted tone, we opt for using the MelGAN discriminator as our baseline. This choice is based on its superior performance in handling heavy distortion settings, as demonstrated the experiments of Wright \emph{et al.} \cite{wright2023adversarial}, viewing the MelGAN discriminator used by them as a variant of MSD. }
Table \ref{experiment_result} shows that the proposed GAN-based approach (i.e., MSD$+$MPD) consistently outperforms the existing GAN-based approach (i.e., MSD) across all the three objective metrics. 
% We can consider the MelGAN discriminator used by \cite{wright2023adversarial} as a variant of MSD. 
Notably, when MPD and MSD were applied in GAN training, these discriminators, originally designed for capturing the diverse periodic patterns, helped the generator produce a more realistic waveform. 
Although our GAN-based model does not include any spectral-based discriminators, for $L1_{mel}$, our method shows a slight improvement on the low-gain tones from EGDB. In general, this result suggests that the combination of MSD and MPD leads to better VA modeling than MSD alone.

Table \ref{experiment_result} also shows that, compared to the supervised baseline~\cite{8682805}, 
the proposed GAN-based approach (MSD$+$MPD) does not lead to better results in ESR. However, for the challenging case of the high-gain tone BD-2 from the EGFxset, the proposed GAN-based approach outperforms the supervised baseline~\cite{8682805} greatly in all the three metrics, especially for $L1_{mel}$ and FAD. 
Informal listening to the generation result (examples available on the demo page) also shows that the proposed model performs perceptually better.\footnote{While we have not conducted a formal subjective evaluation of the implemented models, the listening test reported by the prior work of Wright \emph{et al.}~\cite{wright2023adversarial} has suggested that their GAN-based model (i.e., using MSD alone for the discriminator)  outscores supervised models in human evaluation.} Using MSD alone also outperforms the supervised baseline in $L1_{mel}$ and FAD.
Together, this suggests the advantage of the GAN-based loss for high-gain tones.

%and also demonstrates significant improvement on the high-gain dataset compared to the supervised baseline.
%Comparing our method with the supervised baseline, we find that our method  

\subsection{Clean Audio Combination}

Next, we consider the case where we use the clean audio from both dataset as the input to our generator, no matter whether the target tone is from EGDB or EGFxset. 
Table~\ref{experiment_result} shows that applying this ``both'' input data setting can further boost the performance of all metrics using a paired dataset in a high-gain BD-2 target tone. In other words, the incorporation of EGDB clean tones contributes positively to the modeling of the target tone from EGFxset. This result provides more empirical evidences of the advantage of the GAN-based approach.

\woosungvised{
Interestingly, in contrast, we observe that incorporating clean data from EGFxset does not significantly contribute to modeling any of the target tones from EGDB. 
%Interestingly, in contrast, integrating clean data from EGFxset seem not aid in modeling any of the target tones from EGDB. 
We conjecture that this is due to the differences in music content between the two datasets (e.g., licks versus single notes).
%EGFxset alone does not have note sequences and transitions between individual notes, so adding EGDB inputs can be beneficial. 
Since EGFxset does not have note sequences and transitions between individual notes, adding inputs from EGDB could offer advantages. 
% But, this is not the case the other way around.
However, this is not the case when considering the reverse scenario.
% However, the opposite case is not necessarily helpful because adding clean data from EGFxset  does not make the model's performance significantly.
% However, it is not the case for the the other way around.
%incorporating content from EGDB, which includes , could help the generator react more precisely to clean audio input.
}

\section{Discussion}

\subsection{Benefit of the GAN-based Approach for VA Modeling}
% Here we are interested in one question : Is an unsupversied learning really a necessary approach to an amplifier modeling task? For validate this question, we choose one of the worst cases from the testing set that the supervised approach not performing well to make an case analysis. As one of our ultimate goal is to find a way that performs better than the relatively traditional approach. 
Due to the limits in the amount of available paired data and computation resource, we have only considered two datasets and four tones in total in our experiments. 
However, the experimental result has already revealed potentials of the GAN-based approach over the prevailing supervised approach for VA modeling. To illustrate this point further, we consider the tone that the supervised baseline~\cite{8682805} does not perform well, namely, the EGFxset BD-2, and conduct a case analysis.

%Here, we are interested in answering one question: Is unsupervised learning truly a useful approach for an amplifier modeling task? As one of our ultimate goals is to identify an approach that outperforms traditional methods.
%To validate this question, we choose one of the worst cases from the testing set where the supervised approach does not perform well, and conduct a case analysis.

% In order to see the difference between a supervised approach and our approach. We sampled one case in EGfxset Bd-2 testing set to see its difference from spectral aspect. As shown in Figure~\ref{comparison_figures}, 
% We plot these audios to a mel-spectorgram for a easy observation of the frequency and it's harmonic. The high frequency of the audio from supervised approach are missing their tail compared to the target. On the contrary, high frequency of the audio from MSD$+$MPD can retail through the time axis. 
%To illustrate the disparity between a supervised approach and our method, we sampled one case from the EGfxset BD-2 testing set and compared the spectral aspects. As depicted in 

Figure~\ref{comparison_figures} plots the mel-spectrogram of a sampled target audio rendered with the BD-2 tone from EGFxset, along with the mel-spectrograms of the generation result of the supervised baseline~\cite{8682805} and the proposed model given the corresponding clean signal.
%we plotted these audios as for ease of observing frequency and its harmonics. It is evident 
We can see that many high-frequency harmonics are missing in the result of the supervised baseline. In contrast, they are \woosungvised{effectively}
%nicely
captured by the proposed model throughout the time axis. \ericrevised{Similar observations can be found for other samples for the BD-2 tone.}

From the viewpoint of digital signal processing (DSP), higher gain value implies more non-harmonic high-frequencies in the audio. 
While such non-harmonic high-frequencies may not be well captured by supervised loss functions such as the ESR, they can be better dealt with by adversarial losses such as MSD.  
Compared to MSD only, MSD$+$MPD can perform even better for such high-gain tones.  
We speculate that adding MPD helps, for MPD operates directly on equally-spaced sample points of the audio waveform in its original temporal resolution (e.g., 44.1\,kHz),  while MSD involves downsampling operations of the waveform. The downsampling operations of MSD may have limited its strength in assessing high-frequency components.

Informal listening also shows that the supervised baseline can already model the tones in EGDB well, and that for these tones the GAN-based models do not offer obvious advantages. MSD$+$MPD only performs better on ESR for the Marshall tone. However, we note that the GAN-based models are trained under an unpaired setting, which makes it easier to scale up the training data. Future work can further exploit this advantage by applying data augmentation methods or devising more advanced training framework.

\subsection{Artifacts Generated by the Proposed Model
}
During case analysis, we found that the proposed model is still not perfect and there is room for improvement. In particular, the audio generated by the proposed model may exhibit some artifacts. From Figure~\ref{comparison_figures_2}, MSD$+$MPD generate a splash of harmonics that does not exist in the target mel-spectrogram. It is unclear why the combination of MSD and MPD discriminators cannot detect such artifacts and accordingly prevent the generator from generating them. 
However, as both MSD and MPD are discriminators operating directly on the audio waveforms, it might be interesting to incorporate spectral-based discriminators that operate on time-frequency representations to seek possible improvement. For example, in neural audio compression task~\cite{kumar2024high}, researchers have shown that splitting the STFT into multi sub-bands allows each sub-discriminator focus on specific frequency bands, providing stronger gradient signals to the generator.

Another key observation is that, while both the supervised baseline model and the proposed MSD$+$MPD model can result in low ESR values, both of them do not yet model the high-frequency components perfectly, especially for the sustain of notes.
This suggests that ESR may not be good enough either as a training objective or an objective evaluation metric for amplifier modeling. Future work can explore other auxiliary losses either for the supervised approach or the GAN-based approach.
%, as mentioned in Section~\ref{mel_spec_loss}.  Adding  to discriminator can be consider in the future work.

\section{Conclusion}

In this paper, we have proposed a new GAN-based model for VA modeling by incorporating the MPD discriminator developed in research on neural vocoders.
With experiments on two datasets, we showed that the new model leads to improvement across a range of objective metrics over existing supervised and GAN-based models. Moreover, we demonstrated the benefit of a new scenario where combining clean audio from different datasets enhances GAN training, leading to further performance improvement. %Although our results exhibit comparable performance to supervised approaches, our model still exhibits high-frequency artifacts, which are critical in amplifier-rendered audio. Furthermore, challenges such as longer training times and instability in the training process persist in our unsupervised approach.
Following this light, future work can  explore more advanced discriminator architectures to reduce model size, speed up training time, or further reduce artifacts. %Moreover, additionally, investigating techniques to improve the stability of training processes could also prove valuable.
It would also be interesting to apply the GAN-based approach to datasets with greater diversity in musical content, guitar tone, and recording conditions.

\section{ACKNOWLEDGEMENTS}
The authors would like to thank the support from the Featured Area Research Center Program within the framework of the Higher Education Sprout Project by the Ministry of Education of Taiwan (113L900901 /113L900902 /113L900903).

\nocite{*}
\bibliographystyle{IEEEbib}
\bibliography{DAFX_24_camera_ready} % requires file DAFx24_tmpl.bib

\end{document}